\documentclass[%
 reprint,
 amsmath,amssymb,
 aps,
]{revtex4-2}

\usepackage{graphicx}
\usepackage{dcolumn}
\usepackage{bm}

\def\e{\begin{equation}}
\def\f{\end{equation}}
\def\_#1{{\bf #1}}

\def\.{\cdot}
\def\x{\times}

\usepackage{esvect}
\usepackage[utf8]{inputenc}
\usepackage[T1]{fontenc}
\newcommand*{\affaddr}[1]{#1}
\newcommand*{\affmark}[1][*]{\textsuperscript{#1}}
\begin{document}

\preprint{APS/123-QED}

\title{Chiral Surface Wave propagation with Anomalous Spin-momentum Locking}

\author{%
Sara M. Kandil\affmark[1], Dia’aaldin J. Bisharat\affmark[2] and Daniel F. Sievenpiper\affmark[1]\\
\affaddr{\affmark[1]\textit{University of California San Diego, Department of Electrical and Computer Engineering \\
 La Jolla, CA 92093, U.S.A}}\\
\affaddr{\affmark[2]\textit{City University of New York, Graduate Center\\ New York, New York 10016, USA}}\\
}






\begin{abstract}
The ability to control the directionality of surface waves by manipulating its polarization has been of great significance for applications in spintronics and polarization-based optics. Surface waves with evanescent tails are found to possess an inherent in-plane transverse spin which is dependent on the propagation direction while an out-of-plane transverse spin does not naturally occur for surface waves and requires a specific surface design.  Here, we introduce a new type of surface waves called chiral surface wave which has two transverse spins, an in-plane one that is inherent to any surface wave and an out-of-plane spin which is enforced by the design due to strong $x$-to-$y$ coupling and broken rotational symmetry. The two transverse spins are locked to the momentum. Our study opens a new direction for metasurface designs with enhanced and controlled spin-orbit interaction by adding an extra degree of freedom to control the propagation direction as well as the transverse spin of surface waves.


\end{abstract}
\maketitle
\section{\label{sec:level1}Introduction}
Spin is a universal property inherent to electrons and photons. Electron spin has been the origin of many interesting phenomena such as Spin-Hall Effect (SHE) in which electrons with opposite spins propagate in opposite directions \cite{hirsch1999spin,hasan2010colloquium}. This has opened the door for many applications in spintronics and quantum physics \cite{ling2017recent,wolf2001spintronics}. On the other hand, a photon's spin is associated with its polarization state, described as the handedness of the circular polarization of the electric and/or magnetic fields. And despite electrons and photons being fundamentally different particles, they reveal similar spin-related properties among which is the SHE. It was recently discovered that analogous to SHE in electrons, surface waves (SWs) with evanescent tails obtain an in-plane transverse spin that is locked to the propagation direction \cite{bliokh2014extraordinary,bliokh2012transverse, bliokh2015quantum}. This is also known as spin-momentum locking which is defined as the right-hand triplet formed of the decay, spin and propagation constant \cite{van2016universal,bliokh2015spin}. During the past decade, several photonic platforms were shown to be able to control the propagation direction of light through altering its T-spin \cite{o2014spin, kandil2021c, dia2017guiding,lin2013polarization,mazor2020routing,huang2013helicity}. 



This spin-dependent directionality of surface waves originates from its spin-orbit coupling and can occur on planar metal surfaces without any structures \cite{rodriguez2013near}. However, it is believed to be very small and not easily observed unless enhanced by breaking the spatial inversion symmetry of the surface. This has been studied in the literature in various ways including designing metasurfaces or waveguides with engineered anisotropy \cite{o2014spin, kandil2021c, dia2017guiding,lin2013polarization,mazor2020routing}, gradient metasurface designs \cite{huang2013helicity} and bandgap materials \cite{ruan2020analysis,bisharat2019electromagnetic} or through inversion symmetry breaking by the near-field interference of a dipole source near the surface \cite{picardi2017unidirectional,picardi2018janus}. The ability to control the directionality and polarization of SWs by engineering the metasurface designs is pivotal for many applications in valleytronics \cite{schaibley2016valleytronics} and polarization-based optics such as beam splitting \cite{yoon2018geometric} and spin-based waveguiding \cite{lefier2015unidirectional,kandil2021c}. It was proved in \cite{bliokh2014extraordinary} that any SW obtain an in-plane T-spin. However, the out-of-plane T-spin does not naturally occur for SWs and need to be extrinsically enforced by the design \cite{mazor2020routing}.

In this paper we introduce a new type of SW called chiral SW (CSW). CSW is defined as a SW that possesses two T-spins, an in-plane T-spin which is inherent to any SW and an out-of-plane T-spin which is formed due to the in-plane field rotation that is enforced by the metasurface design. We study analytically and experimentally an L-shaped metasurface design that supports CSW. We show that the coupling between two orthogonal modes each of which has in-plane T-spin can result in forming a new mode that has an out-of-plane T-spin as well as the in-plane T-spin. Both T-spins are locked to the propagation direction of the SW due to the broken rotational symmetry of the L-shape design resulting in spin-dependent directional propagation. This work provides a platform with an extra degree of freedom for controlling the SOI of SWs by adding a new T-spin. It also means that the CSW can be excited using a CP source in the transverse plane ($xz$) as well as in the same plane as the surface ($xy$), which is more practical to demonstrate in experiments.


\section{L-shape Metasurface Design}
\begin{figure}[h!]
	
	\hspace*{-0.1cm}\includegraphics[width=8.5cm,height=5cm]{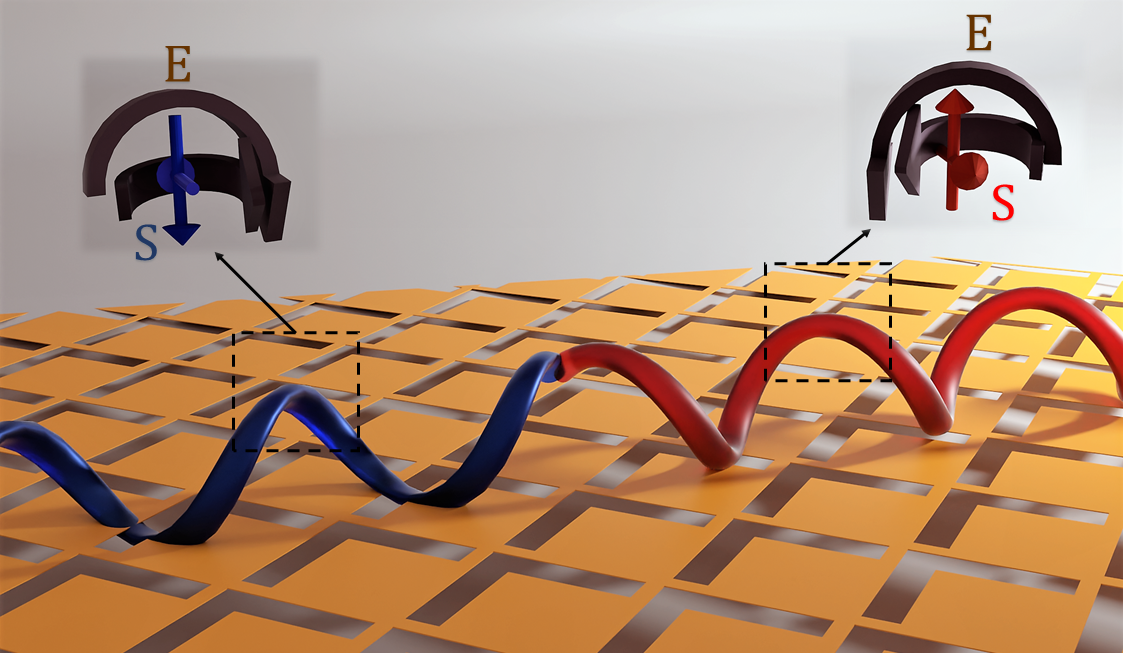}
	\caption{\label{fig:fig1} Schematic representation of the CSW propagation on the L-shape metasurface showing the two T-spins which both flip sign when the propagation direction is reversed.}
\end{figure}
Fig.~\ref{fig:fig1} shows a schematic representation of the CSW propagation on the L-shaped metasurface when excited with a linearly polarized source at the center. As depicted, the CSW possesses two T-spins due to in-plane and out-of-plane E-field rotations. The SW propagates along the diagonal of the metasurface where the two spins flip when the momentum is reversed demonstrating the spin-momentum locking feature.

The significance of CSWs is that unlike other SWs, their T-spin direction is enforced and controlled by the design and is not intrinsic to the wave itself which provides an additional degree of freedom to tailor it. In this section, we explain in terms of analyzing the L-shape design, the design characteristics that can support CSWs. 
\subsection{\label{sec:2level1} Strong x-to-y Coupling}

Chirality of an object is defined as being not superimposable with its mirror image \cite{mun2020electromagnetic,caloz2020electromagnetic}. On the other hand, the chirality of the wave is defined as having rotating electric or magnetic fields (RCP or LCP) while propagating. The value and direction of the chirality of light can be calculated using the vector spin density, S, normalized per one photon in units $\hbar$ \cite{bliokh2014extraordinary,yermakov2016spin} expressed as:
\begin{eqnarray}
\label{eq:one}
\textbf S = \textrm{Im} \left\{ \frac{\textbf E^*\x \textbf E+ \textbf H^*\x \textbf H}{|\textbf{E}|^2+|\textbf{H}|^2} \right\}=\textbf S_\textrm{e}+\textbf S_\textrm{m}
\end{eqnarray}
where $\textrm{S}_\textrm{e}$ and $\textrm{S}_\textrm{m}$ denote the electric and magnetic components of the spin. SWs that are TE or TM are proved to already possess out-of-plane field rotations resulting in an in-plane T-spin \cite{bliokh2014extraordinary}. For the design to support an additional out-of-plane T-spin, it has to enforce an in-plane E-field or H-field rotation. As we will see in this section, the design does not have to be chiral to support this chiral mode but it has to have strong $x$-to-$y$ coupling, which refers to the ability of the design to induce a y-polarized E-field when excited with an x-polarized dipole source and vice-versa.
	

\begin{figure}[h!]
	
	\hspace*{-0.1cm}\includegraphics[width=9cm]{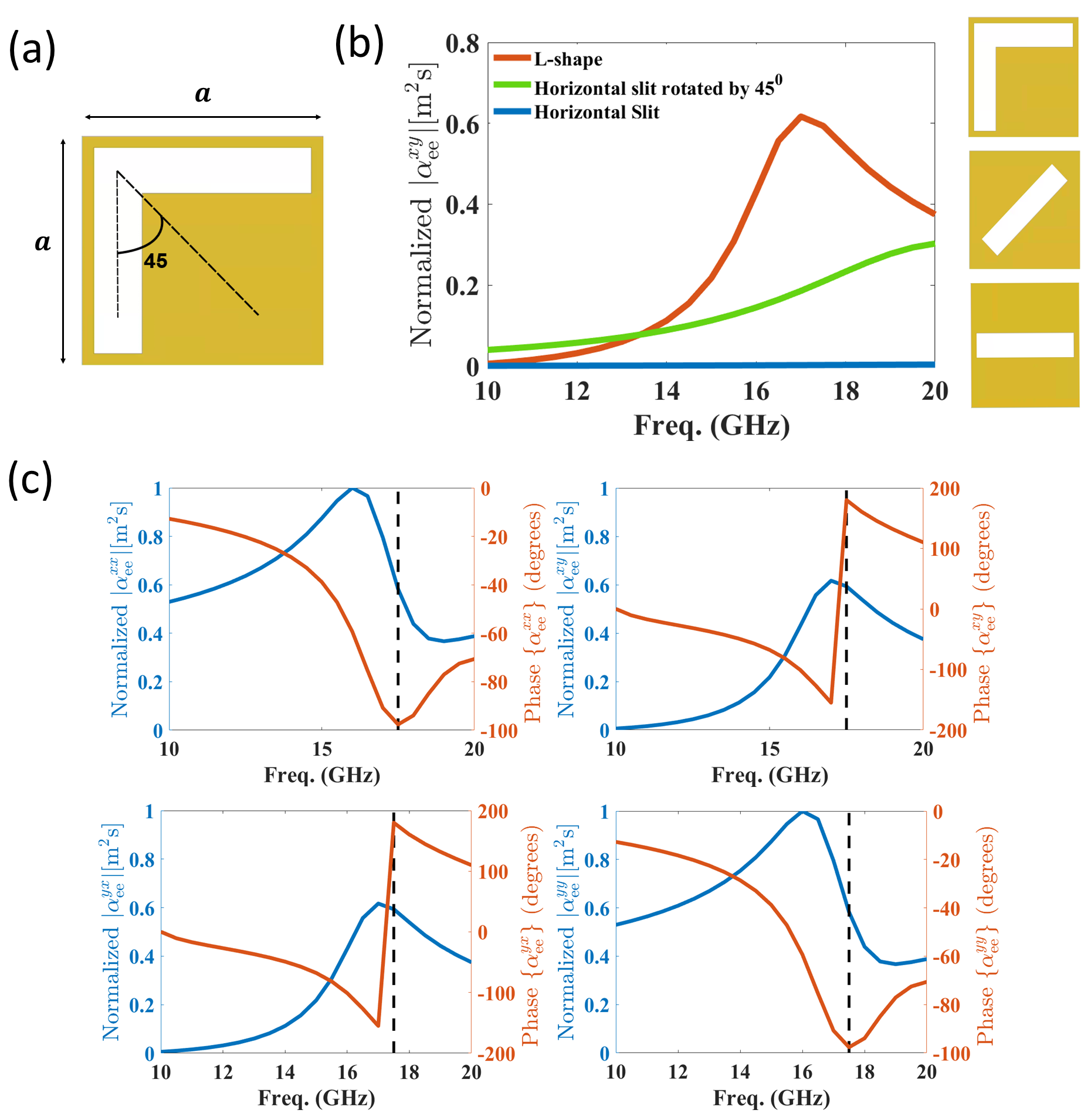}
	\caption{\label{fig:fig2} (a) Schematic of the L-shape unit cell showing its $45^0$ mirror symmetry. (b) Comparison of the extracted normalized magnitude of $\alpha_{ee}^{xy}$ for the L-shape, slit rotated by $45^0$ and horizontal slit. (c) Extracted normalized magnitude and phase of the $\alpha_{ee}$ tensor components of the L-shape. The vertical dotted line highlights the mode frequency 17.5GHz.}
\end{figure}

The L-shape is characterized by two key features that make it ideal for supporting an out-of-plane transverse spin-splitting leading to confined spin-dependent propagation. First it has $\pm 45^0$ mirror symmetry with respect to the $xz$- and $yz$- planes which means it has a non-zero $x$-to-$y$ coupling \cite{achouri2021fundamental} as depicted in the schematic of the L-shape unit cell shown in Fig.~\ref{fig:fig2}(a) where $a$ is the periodicity of the L-shape square unit cell. It is important to note that since the L-shape possesses a mirror symmetry, it is not defined as a chiral structure. The electric $x$-to-$y$ coupling factor can be represented by the electric polarizability, $\alpha_{ee}^{xy}$ (or $\alpha_{ee}^{yx}$), which is a component of the electric polarizability tensor defined in the following equation:
\begin{eqnarray}
\label{eq:two}
\textbf p = \overline{\overline{\alpha}}_\textrm{ee}.\textbf{E}_\textrm{inc}+\overline{\overline{\alpha}}_\textrm{em}.\textbf{H}_\textrm{inc}, 
\end{eqnarray}
where $\bf p$ is the electric dipole moment and $\bf{E}_{\textrm {inc}}$ and $\bf{H}_{\textrm {inc}}$ are the incident electric and magnetic fields, respectively. $\overline{\overline{\alpha}}_\textrm{ee}$ and $\overline{\overline{\alpha}}_\textrm{em}$ are the electric and electro-magnetic polarizability tensors, respectively. For simplicity, we assume we only excite with an electric dipole so only $\overline{\overline{\alpha}}_\textrm{ee}$ will be considered throughout this study. The $\overline{\overline{\alpha}}_\textrm{ee}$ is described as:
\begin{eqnarray}
\label{eq:three}
\overline{\overline{\alpha}}_\textrm{ee}= 
\begin{bmatrix}
\alpha_{ee}^{xx} & \alpha_{ee}^{xy}\\
\alpha_{ee}^{yx} & \alpha_{ee}^{yy}
\end{bmatrix}
\end{eqnarray}

For clarification, Fig.~\ref{fig:fig2}(b) shows a comparison of the normalized magnitude of $\alpha_{ee}^{xy}$ for three different shapes: horizontal slit, slit rotated by $45^0$ from the horizontal axis and the L-shape. The polarizability tensor is calculated using the formulations and code presented in \cite{asadchy2019modular}. As is observed, the three shapes are achiral but the latter two shapes have $45^0$ mirror symmetry and therefore their $\alpha_{ee}^{xy}$ is non-zero. A non-zero $x$-to-$y$ coupling is essential for the shape to support in-plane E-field rotation as will be derived here. The dimensions of the L-shape used in this study are: $a=5mm$, length and width of the slits are 4.5 mm and 1 mm, respectively. The $|\alpha_{ee}^{xy}|$ of the L-shape has a peak value around 17.5 GHz which is the resonance frequency where the chiral surface mode is found to be excited. 

Fig.~\ref{fig:fig2}(c) shows the extracted magnitude and phase of the components of ${\alpha}_\textrm{ee}$ tensor of the L-shape. It can be observed that at the resonance frequency of the L-shape, 17.5 GHz, the four tensor components have equal normalized magnitude of 0.6 while the phases are $-90^0$ for $\alpha_{ee}^{xx}$ and $\alpha_{ee}^{yy}$ and $180^0$ for $\alpha_{ee}^{xy}$ and $\alpha_{ee}^{yx}$. The magnitudes and phases of $\overline{\overline{\alpha}}_\textrm{ee}$ can be substituted in Eq.~\ref{eq:three} giving the following matrix:
\begin{eqnarray}
\label{eq:four}
\overline{\overline{\alpha}}_\textrm{ee}= 
\begin{bmatrix}
-i0.6 & -0.6\\
-0.6 & -i0.6
\end{bmatrix}
\end{eqnarray}
By substituting Eq.~\ref{eq:four} in Eq.~\ref{eq:two}, the electric dipole moment can be written as: 
\begin{eqnarray}
\label{eq:five}
\textbf p \propto -\textrm E_\textrm{inc}^{x} \left (i0.6 \widehat{\textbf{e}}_\textrm{x}+0.6 \widehat{\textbf{e}}_\textrm{y}\right)- \textrm E_\textrm{inc}^{y}\left (0.6 \widehat{\textbf{e}}_\textrm{x}+i0.6 \widehat{\textbf{e}}_\textrm{y}\right).
\end{eqnarray}
From Eq.~\ref{eq:five}, one can conclude that an in-plane E-field rotation is supported by the L-shape when excited with x-polarized ($\textrm E_\textrm{inc}^{y}=0$)or y-polarized ($\textrm E_\textrm{inc}^{x}=0$) dipole source. Hence, the following condition can be derived for the metasurface design to induce an in-plane circularly polarized mode:
\begin{eqnarray}
\label{eq:six}
\alpha_{ee}^{xx}=\pm i\alpha_{ee}^{xy}.
\end{eqnarray}
\subsection{Broken Rotational Symmetry and Spin-splitting}
\begin{figure}[h!]
	
	\hspace*{-0.1cm}\includegraphics[width=9.7cm,height=10cm]{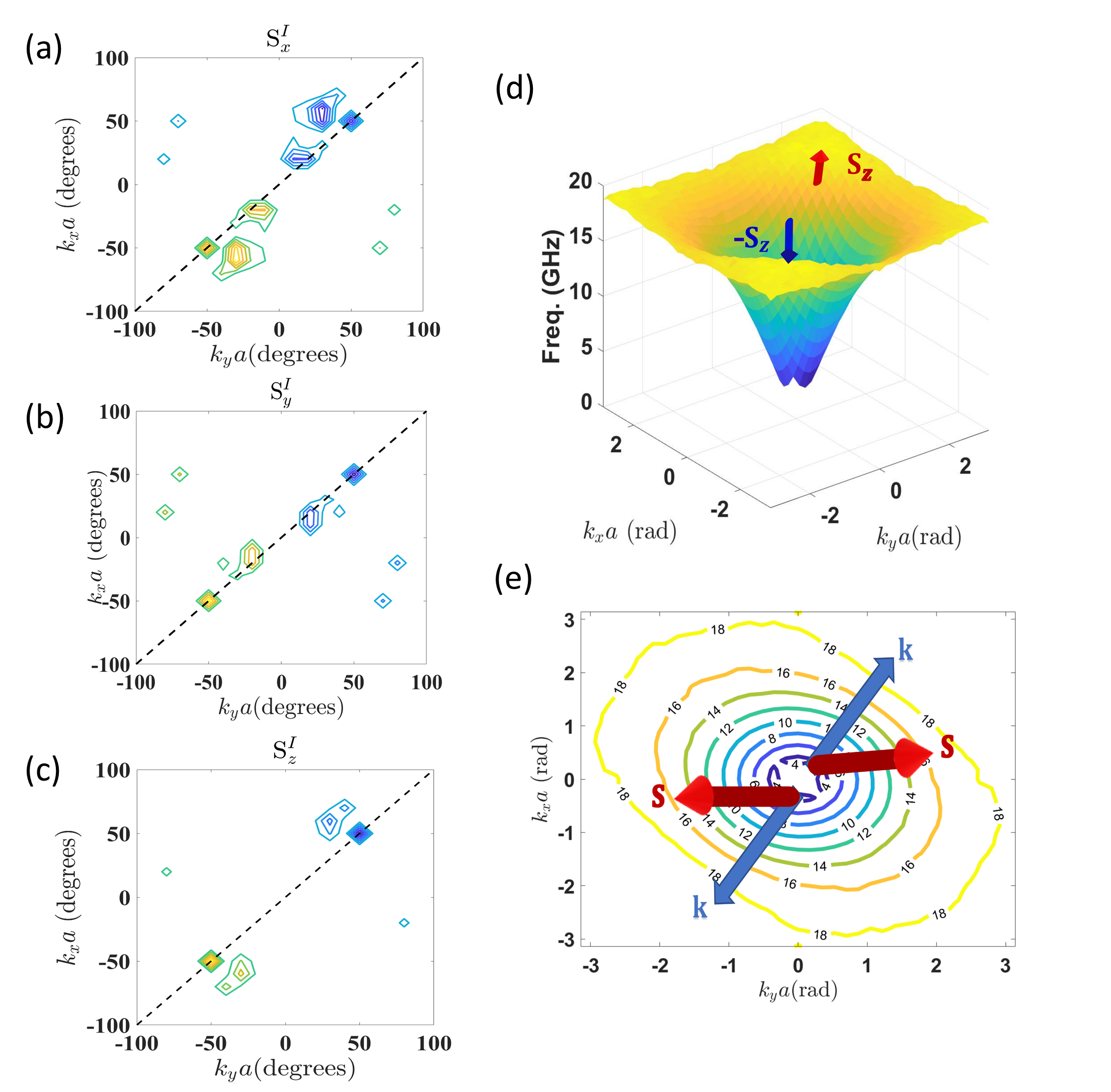}
	\caption{\label{fig:fig3} (a) Numerically calculated integrated spin number k-space maps in x (a),y (b) and z directions (c). (d) The numerically calculated 3D dispersion diagram of the L-shape and (e) its EFC using eigenmode simulation. The directions of T-spins and the k-vector are presented showing the spin-momentum locking feature.}
\end{figure}
Secondly, the L-shape has broken rotational symmetry. It was shown in the literature that the broken rotational symmetry of the metasurface can result in spin-dependent splitting in the momentum space by inducing an additional geometric phase for LCP incidence with respect to RCP incidence \cite{liu2015photonic,huang2013helicity,bliokh2008coriolis}. In Fig.~\ref{fig:fig3}(a), we show the spin-splitting in k-space along x,y and z by calculating the integral spin density, $S^I$, defined as:
\begin{eqnarray}
\label{eq:seven}
S^I_x(k_x,k_y)=\int_{x=0}^{x=a}\int_{y=0}^{y=a}S_x(x,y,k_x,k_y) \:ds.
\end{eqnarray}
\begin{eqnarray}
\label{eq:seven}
S^I_y(k_x,k_y)=\int_{x=0}^{x=a}\int_{y=0}^{y=a}S_y(x,y,k_x,k_y) \:ds.
\end{eqnarray}
\begin{eqnarray}
\label{eq:seven}
S^I_z(k_x,k_y)=\int_{x=0}^{x=a}\int_{y=0}^{y=a}S_z(x,y,k_x,k_y) \:ds.
\end{eqnarray}
where $S_x$, $S_y$ and $S_z$ are x, y and z components of the spin density calculated using Eq.~(\ref{eq:one}). The integral spin densities are evaluated numerically using the eigenmode solver by integrating over the surface area of the L-shape unit cell at different points in the 2D k-space $(k_xa,k_ya)$ to generate the spin maps presented in Fig.~\ref{fig:fig3}(a), (b) and (c). The $S^I_x$ and $S^I_y$ represent the in-plane transverse spins while $S^I_z$ represents the out-of-plane transverse spin. The in-plane as well as the out-of-plane spins splitting can be observed where the orange contours indicate positive $\rm S^I$ (spin up) while the blue ones indicate a negative $\rm S^I$ (spin down). It can be shown that the spins split along the diagonal plotted as a dotted line at the same k-coordinates of $(-50^0,-50^0)$ (spin-up) and $(50^0,50^0)$ (spin-down). This means that the two transverse spins will be supported by the SW having a propagation constant matching these k-coordinates. The diagonal line here corresponds to the direction of the propagation of the chiral mode $(\widehat{\textbf{e}}_\textrm{x}+\widehat{\textbf{e}}_\textrm{y})$ where the spin-momentum locking takes place. The spin-splitting in k-space results in spin-dependent propagation in real space as will be shown in Sec.~\ref{sec:level2}.

\subsection{Self-Collimation of L-shape}
The L-shape metasurface is characterized by having high self-collimation. Self-collimation is a property intensely studied for photonic crystals and it arises due to special dispersion characteristics in which the wave is forced to propagate in a confined direction without spreading \cite{prather2007self}. This is usually represented by a flat equifrequency contour (EFC). Breaking the rotational symmetry of the lattice unit cell has shown to present new features in engineering the dispersion diagram among which is the high self-collimation and tilted EFCs \cite{kurt2012crescent}. This can be shown for the L-shape in the flat EFC presented in Fig.~\ref{fig:fig3}(e). The EFC can be obtained by slicing the 3D dispersion diagram shown in Fig.~\ref{fig:fig3}(d) at different frequencies. Both are calculated numerically using eigenmode simulations. The poynting vector of the SW can have different directions depending on the curvature of the EFC. The flatter the contour, the higher the self-collimation is since the wave is then forced to propagate in single direction, that is normal to the flat contour. The highly self-collimated beams have found to be useful for making compact easily integrated optical chips for strongly guiding waves in the presence of bends and defects \cite{gutierrez2019independent}. As shown in Fig.~\ref{fig:fig3}(b), the EFCs of the L-shape are found to be tilted at an angle of $45^0$ from the y-axis verifying the k-vector direction along $(\widehat{\textbf{e}}_\textrm{x}+\widehat{\textbf{e}}_\textrm{y})$ as predicted earlier. 
\section{\label{sec:level2} Numerical Simulation Results}
\begin{figure}
	
	\hspace*{-0.1cm}\includegraphics[height=12cm]{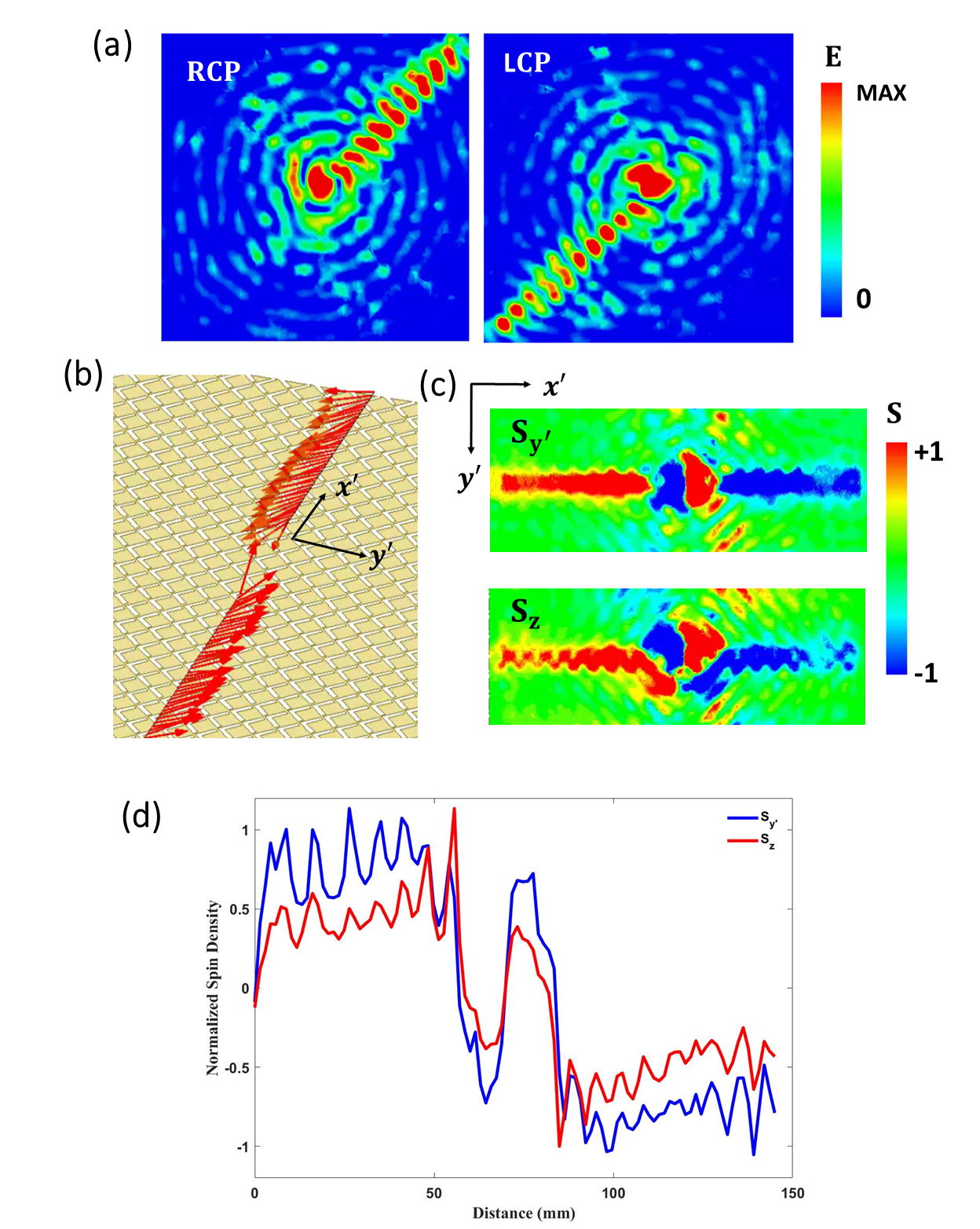}
	\caption{\label{fig:fig5} (a) Magnitude of E-field map of the L-shape metasurface at 17.5GHz GHz when excited with an $\textbf{ E}\textsubscript{x}+i\textbf{E}\textsubscript{y}$ dipole source (left) and $\textbf{ E}\textsubscript{x}-i\textbf{E}\textsubscript{y}$ dipole source (right). For simplicity, the x-y coordinates are rotated 45 degrees and denotes as x' and y'. Spin density study of the chiral wave when excited with $\textrm{E}_\textrm{x}$: (b) The calculated vector spin density showing that the direction of spin flips as the propagation direction is reversed. (c) Normalized $S\textsubscript{y'}$ and $S\textsubscript{z}$ 2D maps and (d) normalized T-spins calculated  plotted at a line along the diagonal of the metasurface.} 
\end{figure}

The L-shape metasurface is numerically simulated using Ansys HFSS. Fig.~\ref{fig:fig5}(a) shows the simulated E-field distribution at 17.5GHz where the surface is excited with an in-plane CP source at the center described as $\textbf{E}\textsubscript{x}\pm i\textbf{E}\textsubscript{y}$. The polarization-dependent unidirectional propagation feature of the CSW is depicted where an RCP source excites a surface mode that propagates only forward along the diagonal (left) while an LCP source excites a surface mode propagating only backward (right). The spin density is calculated using Eq.~(\ref{eq:one}) for the CSW and is presented in Figs.~\ref{fig:fig5}(b), ~\ref{fig:fig5}(c) and ~\ref{fig:fig5}(d). For simplicity, the $xy$ axes are rotated $45^0$ in the clockwise direction in which the modified axes are $x'$ and $y'$. 

A vector representation of the spin density is shown in Fig.~\ref{fig:fig5}(b), calculated when the surface is excited with a x-polarized dipole source at the center of the surface. The two induced T-spins are then along y' and z axis. The normalized magnitudes of $S\textsubscript{y'}$ (in-plane T-spin) and $S\textsubscript{z}$ (out-of-plane T-spin) are shown as color maps in Fig.~\ref{fig:fig5}(c) and along a line across the diagonal in Fig.~\ref{fig:fig5}(d). The spin-momentum locking feature is clearly observed from the numerically calculated spin density of the CSW for the two T-spins. This also means an additional degree of freedom for excitation of the CSW which can be directionally excited by an in-plane CP source as well as an out-of-plane CP source.
\section{Experimental Results}
\begin{figure}
	
	\hspace*{-0.2cm}\includegraphics[width=9cm,height=7cm]{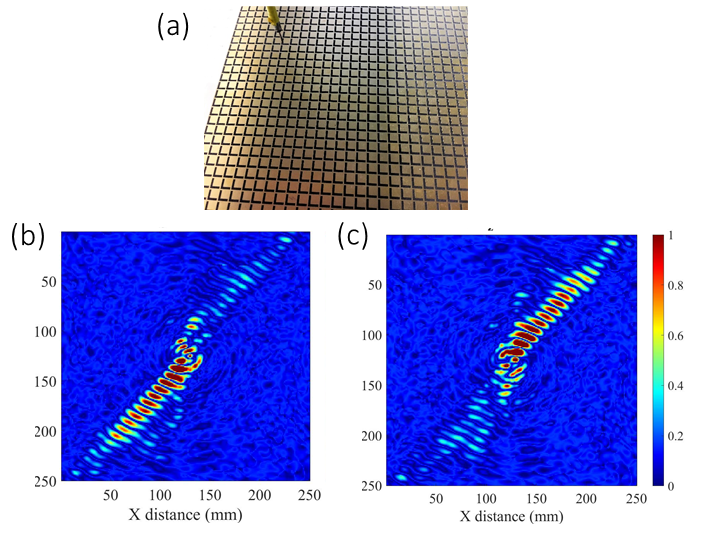}
	\caption{\label{fig:fig6} Measurement results of the L-shape metasurface (a) A photo showing the measurement setup where a probe antenna oriented along the z-axis is used to measure the E-field distribution. An $x$ and $y$ probes were used for the excitation. Magnitude of the $\textrm{E}_\textrm{z}$ distribution measured at 14.5 GHz when excited at the center with an equivalent excitation of (b) $\textbf{E}\textsubscript{x} - i\textbf{E}\textsubscript{y}$ and (c) $\textbf{E}\textsubscript{x} + i\textbf{E}\textsubscript{y}$} 
\end{figure}

The measured results of the L-shape metasurface is presented in Fig.~\ref{fig:fig6}. The L-shape surface is fabricated on a printed circuit board of Rogers 5880 ($\epsilon_r=2.2$) which results in a shift in the frequency of the supported CSW from 17.5 GHz to 14.5 GHz. The magnitude of the $E\textsubscript{z}$ component is measured showing the directional excitation as shown in Fig.~\ref{fig:fig6}(b) and ~\ref{fig:fig6}(c), respectively. The surface is excited with two probe antennae oriented along $x$ and $y$ directions where the magnitude and phase of $E\textsubscript{z}$ are measured for each probe excitation separately. The measured magnitudes and phases are then added with a phase shift of 90 degrees to form an excitation of $\textbf{E}\textsubscript{x}\pm i\textbf{E}\textsubscript{y}$ as depicted in the following equation:
\begin{equation}
\textbf{E}_\textsubscript{z}^{x} \pm i\textbf{E}_\textsubscript{z}^{y}= |\textbf{E}_\textsubscript{z}^{x}|\cos\left(\phi_\textsubscript{z}^{x}\right) +|\textbf{E}_\textsubscript{z}^{y}|\cos\left(\phi_\textsubscript{z}^{y}\pm 90^0\right),
\end{equation}
where $\textbf{E}_\textsubscript{z}^{x(y)}$ is the measured $\textbf{E}_\textsubscript{z}$ when excited with a probe along $x (y)$-axis. Using the $45^0$ mirror symmetry of the L-shape surface described in Sec.~\ref{sec:2level1}, an excitation with an x-probe along the vertical arm of the L-shape can be mirrored to a y-probe excitation along its horizontal arm. Since the excited mode is propagating along the diagonal of the surface, $\textbf{E}_\textsubscript{z}^{y}$ can be extracted from $\textbf{E}_\textsubscript{z}^{x}$ by flipping the vector across the normal to the diagonal of the surface or by simply rotating the measured E-field vector by 180 degrees. 
\section{Conclusion}
In this work, we have shown numerically and experimentally the possibility of the excitation of a new type of surface waves called chiral surface wave which unlike other surface waves, possesses two transverse spins, an in-plane transverse that is inherent to any surface wave and an out-of-plane transverse spin that is enforced by the L-shape design. We showed that due to the $45^0$ mirror symmetry and the broken rotational symmetry of the L-shape design, it results in strong in-plane E-field rotation and spin splitting in k-space. The two transverse spins follow the spin-momentum locking law resulting in spin-dependent directional propagation. Another important feature of the L-shape design is that it has flat EFC which results in a highly collimated surface wave. This work provides a new degree of freedom for controlling the spin-orbit interaction of evanescent waves by adding an extra transverse spin. We show that supporting a highly collimated spin-dependent unidirectional modes with strong spin-orbit interaction can be done using a homogeneous metasurface design without the complexity of designing bandgap materials or interfaces between different materials.

This work is supported by the AFOSR under Grant No. FA9550-16-1-0093 as well as the ONR under Grant No. N00014-20-1-2710.
\nocite{*}
\section*{References}
\bibliography{apssamp}
\end{document}